\documentstyle[epsfig,12pt,a4p,subfigure]{article}

\newcommand{\be}{\begin{equation}}
\newcommand{\ee}{\end{equation}}
\newcommand{\bea}{\begin{eqnarray}}
\newcommand{\eea}{\end{eqnarray}}
\newcommand{\bean}{\begin{eqnarray*}}
\newcommand{\eean}{\end{eqnarray*}}

\newcommand{\gapproxeq}{\lower
.7ex\hbox{$\;\stackrel{\textstyle >}{\sim}\;$}}
\newcommand{\lapproxeq}{\lower
.7ex\hbox{$\;\stackrel{\textstyle <}{\sim}\;$}}

\newcommand{\pipi}{\mbox{$\pi^{+}\pi^{-}$} }

\begin{document}
\begin{titlepage}
\begin{tabbing}
wwwwwwwwwwwwwwwright hand corner using tabbing so it looks neat and in \=
\kill
% \> {hep-ph/01xxxxx}   \\
\> {OUTP-0114P}   \\
% \> {BHAM-HEP/00-xx}   \\
\> {15 March 2001}
\end{tabbing}
\baselineskip=18pt
\vskip 0.7in
\begin{center}
{\bf \LARGE Scalar Glueball-$q\bar{q}$ Mixing above 1 GeV and implications
for
Lattice QCD}\\
\vspace*{0.9in}
{\large Frank E. Close}\footnote{\tt{e-mail: F.E.Close@rl.ac.uk}} \\
\vspace{.1in}
{\it Dept of Theoretical Physics}\\
{\it University of Oxford}\\
{\it 1 Keble Rd, Oxford, OX1 3NP,UK}\\
%{\it and}\\
%{\it Rutherford Appleton Laboratory}\\
%{\it Chilton, Didcot, OX11 0QX, UK}\\
\vspace{0.1in}
{\large Andrew Kirk}\footnote{\tt{e-mail: ak@hep.ph.bham.ac.uk}} \\
{\it School of Physics and Astronomy}\\
{\it Birmingham University, Birmingham, B15 2TT}\\
%\vspace*{0.1in}
\end{center}
%\maketitle
\begin{abstract}
Lattice QCD predictions have motivated several recent studies of the
mixing between the predicted $J^{PC} = 0^{++}$
glueball and a $q\bar{q}$ nonet in the $1.3 \to 1.7$ GeV region.
 We show that results from
apparently different approaches have some common features,
 explain why this is so and abstract general conclusions.
  We place particular emphasis on the flavour dependent
constraints imposed by decays of the $f_0(1370)$, $ f_0(1500)$
and $f_0(1700)$ to all pairs of pseudoscalar mesons.
 From these results we  identify a systematic correlation
between glueball mass, mixing, and flavour symmetry breaking and conclude
that
the glueball may be rather lighter than some quenched lattice
QCD computations have suggested. We identify experimental
tests that can determine the dynamics of a glueball in this mass
region and discuss quantitatively the feasibility of
decoding glueball-$q\bar{q}$
mixing.
\end{abstract}
\end{titlepage}
\setcounter{page}{2}
\par

\section{Introduction}

The best estimate for the masses of glueballs comes from
lattice gauge theory calculations, which in the quenched approximation show
\cite{re:lgt}
that the lightest glueball has
$J^{PC} = 0^{++}$
and that
its mass should be in the range
$1.45-1.75$ GeV. While the lattice remains immature for predicting glueball
decays,
Amsler and Close \cite{re:AC,AC96} first pointed out that in lattice
inspired models,
such as the flux tube \cite{re:IP}, glueballs will mix strongly
with nearby $q \overline q$ states with the same
$J^{PC}$~\cite{anisovich}. Recent studies on coarse-grained lattices
appear to confirm that there is indeed significant mixing between $G$
and $q\bar{q}$ together with associated mass shifts,
at least in the $J^{PC}$~=~$0^{++}$
sector~\cite{michael}. If these results survive at finer lattice
spacing, the conclusion will be that glueball-flavour mixing
is a controlling feature of the phenomena in the
$\sim 1.3 - 1.7$ GeV
mass region of meson spectroscopy. It is our purpose in
the present paper to build a phenomenological interpretation of the data
based on intuition from lattice QCD, and to identify the data
needed to confirm it.

To help orient readers, we first present an overview of the paper
and its central conclusions.

The first analyses of $G-q\bar{q}$ mixing used the mass matrix with an
assumed $G-q\bar{q}$ mixing strength
\cite{re:AC,AC96,re:CFL,Wein,mix,Genov}.
Such a mixing between a glueball and a $q\bar{q}$ nonet
 will lead to three isoscalar states of the same
$J^{PC}$. Motivated by Lattice QCD, these analyses focussed on the
physical states in the
glueball
mass region - the
$f_0(1370)$ and $f_0(1500)$ and either predicted the existence of
a further $J=0$ state around 1700 MeV \cite{re:AC,AC96} or assumed that
the $f_J(1710)$ was that state \cite{Wein}. The existence of this
scalar meson is now experimentally verified~\cite{PDG2000}.

These papers differed in what they assumed for the mass of the bare
glueball
relative to the $S \equiv s\bar{s}$, which led to some quantitative
differences
in output. Nonetheless, while these papers at first sight differed in
detail,
  their
conclusions share some common robust features. In particular, the flavour
content of
the states is predicted to have the $n \overline n$ and $s \overline s$ in
phase
(SU(3) singlet tendency) for the $f_0(1370)$ and $f_0(1710)$, and
out of phase (octet tendency) for the  $f_0(1500)$. In section 2 we
review these papers and show why their outputs are similar. In particular
these similarities highlight that further constraints are needed if we are
to
establish whether the bare glueball is at the upper \cite{Wein,Genov} or
lower
\cite{re:AC,AC96} end of the $1.45 - 1.75$ GeV range favoured by quenched
lattice QCD, or
even whether the glueball is required~\cite{klempt}.

There are now extensive data on the production and decay
\cite{WASUM} of the above states. These provide further constraints on
the $G-q\bar{q}$
content. Theoretical analysis of decays is under better control than
production and
so we shall discuss the implications of decays first (section 3).

\par

The WA102 collaboration has published~\cite{etaetapap},
for the first time in a single experiment, a complete data set
for the decay branching
ratios of the
$f_0(1370)$, $f_0(1500)$ and
$f_0(1710)$ to all pseudoscalar meson pairs.
Ref.~\cite{mix} and a preliminary letter by us~\cite{ckmix}
have examined the flavour dependence of scalar meson
decay and how these data constrain the flavour and glue mixing of these
scalar states.
 The results here too agree
with the generic structure found in the mass mixing analyses of section 2.
We identify why this is so and assess the implications. A result that
is more general than any specific mixing scheme is that no pair out of
the three $f_0(1370), f_0(1500), f_0(1710)$ can be in the same pure
$q\bar{q}$ nonet; other degrees of freedom are required.

 We shall see that the results are robust. They confirm lattice results
that $G-q\bar{q}$ mixing is (nearly) flavour blind and suggest that the
preferred glueball mass falls into the mass
range, $\frac{M_S+M_N}{2} > M_G \geq M_N$. Then (section 3.2) we will
investigate
the stability against flavour symmetry breaking.
 From these results we shall identify a systematic correlation
between glueball mass, mixing, and flavour symmetry breaking. To choose
among these
results requires further experimental
tests that can determine the dynamics of a glueball in this mass
region. This brings us to section 5 and production dynamics.

Production by $\gamma \gamma$ is potentially the cleanest as this probes
the

$q\bar{q}$ flavours and phases. We advocate serious study of
$\gamma \gamma \to 0^{++}$ as the sharpest arbiter of the wavefunctions,
but we also warn against overly naive interpretation of $\gamma \gamma$
couplings in the $0^{++}$ sector.
The angular and kinematic dependence of $pp \to pp+ 0^{++}$ also shows
distinct differences among the various mesons, but the dynamical origin of
this is still obscure. We note a possible systematic pattern that
correlates
the
$G$ and flavour mixing in our solutions with the distributions observed in
central production. Further ways of separating the $G-q\bar{q}$ content in
the
$0^{++}$ sector are proposed. Ideally $\gamma \gamma$ couplings can
disentangle the amplitudes and this can then be used to decode the dynamics
of central production.

\section{Mass Mixing}

Based upon intuition from lattice QCD, refs.~\cite{re:AC,Wein,FC00}
investigated the mixing between a $J^{PC} = 0^{++}$ glueball, $G$,
 and a $J^{PC} = 0^{++}$ $q\bar{q}$ nonet in its
vicinity. The detailed assumptions of the two approaches differed
but the outputs were remarkably similar in certain features. We shall
first illustrate why this similarity occurs, abstract its general features
and then propose further tests of the general hypothesis.

\par
In the $|G\rangle \equiv |gg\rangle$, $|S\rangle \equiv |s\bar{s}\rangle$,
$|N\rangle \equiv |u\bar{u}+d\bar{d}\rangle/\sqrt{2}$ basis,
the mass matrix describing the $G-q\bar{q}$ mixing
 was written as follows in ref.~\cite{Wein}:
\begin{equation}
M=\left( \begin{array}{ccc}
M_G & f & \sqrt{2}fr\\
f & M_S & 0\\
\sqrt{2}fr & 0 & M_N
\end{array}\right).
\label{eq:d}
\end{equation}
\noindent Here $f \equiv \langle G|M|S\rangle$
and $r \equiv \langle G|M|N\rangle/\sqrt{2}\langle G|M|S \rangle$
are the mixing strengths between the glueball and the quarkonia states.
For a $G-q\bar{q}$ coupling that is flavour blind, $r = 1$.
Lattice QCD~\cite{Wein} finds for $J^{PC} = 0^{++}$ that $r=1.20\pm0.07$.
$M_G$, $M_S$ and $M_N$ represent the masses of the bare states
$|G\rangle$, $|S\rangle$ and $|N\rangle$, respectively.

Refs.~\cite{re:AC,Wein}
%,mix}
assumed that the mixing
is strongest between the glueball and nearest $q \overline q$
neighbours. With the lattice (in the quenched approximation)
predicting the glueball mass to be in the $1.45-1.75$~GeV region, this has
naturally led attention to focus on
 the physical states $|f_0(1710)\rangle$,
$|f_0(1500)\rangle$ and
$|f_0(1370)\rangle$ as the eigenstates of $M$ with the eigenvalues of
$M_{1}$, $M_{2}$ and $M_{3}$, respectively.
%(An alternative picture could
%involve the states $f_0(1500), f_0(1710), f_0(2000)$;
%we do not discuss this in the present
%paper).
The three physical states can be read as~\cite{mix,ckmix}
\begin{equation} \left( \begin{array}{ccc}
|f_0(1710)\rangle\\
|f_0(1500)\rangle\\
|f_0(1370)\rangle
\end{array}\right)
=U\left(\begin{array}{ccc}
|G\rangle\\
|S\rangle\\
|N\rangle
\end{array}\right)
=\left(\begin{array}{ccc}
x_1 & y_1& z_1\\
x_2& y_2& z_2\\
x_3 & y_3 & z_3
\end{array}\right)
\left(\begin{array}{ccc}
|G\rangle\\
|S\rangle\\
|N\rangle
\end{array}\right),
\label{eq:e}
\end{equation}
where
\begin{equation}
U=\left(\begin{array}{ccc}
(M_1-M_S)(M_1-M_N)C_1&
(M_1-M_N)fC_1& \sqrt{2}(M_1-M_S)rfC_1\\
(M_2-M_S)(M_2-M_N)C_2&
(M_2-M_N)fC_2& \sqrt{2}(M_2-M_S)rfC_2\\
(M_3-M_S)(M_3-M_N)C_3&
(M_3-M_N)fC_3& \sqrt{2}(M_3-M_S)rfC_3
\end{array}\right)
\label{eq:f}
\end{equation}
with $C_{i(i=1,~2,~3)}=[(M_i-M_{S})^2(M_i-M_N)^2+
(M_i-M_N)^2f^2+2(M_i-M_S)^2r^2f^2]^{-\frac{1}{2}}$ and
$\Sigma M_{1+2+3} \equiv \Sigma M_{G+S+N}$.

To focus discussion, we first summarise and compare various mixings
that have been proposed in the literature. In the original analysis
of the glueball-$q \bar{q}$ mixing, ref.~\cite{re:AC, AC96} worked
 at leading order
in perturbation and
obtained
\begin{eqnarray}
\label{mixing}
N_G|\Psi_1\rangle = |G\rangle + \xi ( \sqrt{2} r |N\rangle + \omega
|S\rangle) \nonumber \\
N_s|\Psi_2\rangle = |S\rangle - \xi \omega |G\rangle \nonumber
\\
N_n|\Psi_3\rangle = |N\rangle - \xi  \sqrt{2} r |G\rangle
\end{eqnarray}
where the $N_i$ are appropriate normalisation factors, $\omega \equiv
\frac{M_G - M_N}{M_G -M_S}$ and the mixing parameter
$\xi \equiv \frac{f}{M_G - M_N}$. This leading form is strictly
only valid when both $\xi$ and $\xi \omega << 1$.
The $3 \times 3$ matrix,
eqs. (1)-(3) effectively generalised this to
$O(\xi^2)$. The pQCD
analysis of ref.~\cite{re:CFL}
suggested that the $gg \to q\bar{q}$ mixing amplitude manifested
in $\psi \to \gamma R(q\bar{q})$ is  qualitatively
$ \sim O(\alpha_s) \sim 0.5$. While the absolute value of $\xi$ was
not precisely determined, it nonetheless suggested that $O(\xi^2)$ effects
may be significant, as in eqs.~(\ref{eq:e},\ref{eq:f}).
In particular this introduces
$N$ into $\Psi_2$ and $S$ into $\Psi_3$.

Mixing based on lattice glueball masses leads to two classes of solution
of immediate interest:

\noindent (i) $\omega \leq 0$, corresponding to $G$ in the midst
of the nonet~\cite{re:AC}

\noindent (ii) $\omega > 1$, corresponding to $G$ above the
$q\bar{q}$ members of the nonet~\cite{Wein}.

\noindent The model of Genovese~\cite{Genov} is a particular case where
the $G$ and $S$ are degenerate; mathematically his solution is subsumed
in eq.~(\ref{mixing}) when
$\xi \to 0$; $\omega \to \infty$ with $\xi \omega \to 1$.

Weingarten~\cite{Wein}  constructed his
mixing model based on the scenario from
 lattice QCD that the scalar $s\bar{s}$ state,
in the quenched approximation, may lie lower than the scalar glueball
\cite{Wein,lacock96} (thus $\omega > 1$ in the above formalism).
  In their most recent computation of ref. \cite{Wein},
the input ``bare" masses were
$M_N = 1470$ MeV; $M_S = 1514$ MeV; $M_G = 1622$ MeV and the mixing
strength
$f \equiv \xi \times (M_G - M_N) = 64 \pm 13$ MeV, whereby $\xi \sim 0.4
\pm
0.1$.
 The
resulting mixtures, with errors shown in parentheses, are (up to an overall
phase)
\begin{equation}
\begin{array}{c c c c}
&f_{i1}^{(G)} & f_{i2}^{(S)} & f_{i3}^{(N)}\\
f_0(1710) & 0.86(5) & 0.30(5) & 0.41(9)\\
f_0(1500) & -0.13(5) &  0.91(4) & -0.40(11)\\
f_0(1370) & -0.50(12) &  0.29(9) & 0.82(9)\\
\end{array}
\label{eq:1}
\end{equation}

It is instructive to compare this with the assumption of
ref.~\cite{re:AC,AC96}
where, for example,the $G$ lies between
$n\bar{n}$ and $s\bar{s}$ such that the parameter $\omega \sim
-2$. At first sight this would appear to be quite different
to the above, but if for illustration we adopt $\xi
=0.5 \sim \alpha_s$, the resulting mixing amplitudes are

\begin{equation}
\begin{array}{c c c c}
&f_{i1}^{(G)} & f_{i2}^{(S)} & f_{i3}^{(N)} \\
f_0(1710) & 0.60 & 0.76 & 0.22\\
f_0(1500) & -0.61 & 0.61 & -0.43\\
f_0(1370) & -0.50 &  0.13 & 0.86\\
\end{array}
\label{eq:2}
\end{equation}

It is immediately apparent that
the solutions for the lowest mass state in the two schemes
are similar, as are the relative
phases throughout and also the qualitative importance of the $G$ component
in the high
mass state.  Both solutions exhibit destructive interference between
the $N$ and  $S$ flavours for the middle state.

This
parallelism is not a coincidence. The essential dynamical assumption of
ref.~\cite{Wein} and here is that the basic $G-q\bar{q}$ coupling is
(nearly)
flavour symmetric.
A general feature of such a three way mixing is that in the extreme limit
of
infinitely strong mixing the central state would tend towards flavour octet
with the
outer (heaviest and lightest) states being orthogonal mixtures of
glueball and flavour singlet, namely

\begin{equation}
\begin{array}{c c c}
f_0(1710) & \rightarrow & |q\bar{q}(\bf{1})\rangle + |G\rangle     \\
f_0(1500) & \rightarrow &  |q\bar{q}(\bf{8})\rangle + \epsilon |G\rangle
\\
f_0(1370) & \rightarrow & |q\bar{q}(\bf{1})\rangle -|G\rangle   \\
\end{array}
\label{eq:infmix}
\end{equation}
where $\epsilon \sim \xi^{-1} \to 0$.
In effect, in such an extreme case the glueball would have
 leaked away maximally
to the outer states even in the case (ref.~\cite{re:AC,AC96})
where the bare glueball
(zero mixing) was in the middle of the nonet to start with. The leakage
into
the outer states becomes significant once the mixing strength (off diagonal
 term in the mass matrix) becomes comparable to the mass gap between
glueball
and $q\bar{q}$ states (i.e. either $\xi \geq 1$ or $\xi \omega \geq 1$).
 Even in the zero width approximation of ref.~\cite{re:AC,AC96}
this tends to be the case and when one allows for the widths being of
$O(100)$ MeV while the nonet masses and glueball mass are themselves
spread over only
a few hundred MeV, it is apparent that there will be considerable mixing of
 the glueball into the $q\bar{q}$ nonet. The tendency for the $q\bar{q}$
content to separate into two constructive (``singlet tendency") and one
destructive (``octet tendency") happens for even mild mixing; the complete
leakage
of glueball from the latter is only effected as the mixing indeed tends
towards infinity.

If the $G-q\bar{q}$ coupling is flavour dependent, such that (as above)

\begin{equation}
r \equiv \langle G|M|N\rangle/\sqrt{2}\langle G|M|S \rangle \neq 1
\end{equation}
the ``asymptotic maximal mixing" solution will reflect this. Specifically
(up to overall normalisation factors)

\begin{equation}
\begin{array}{c c c}
f_0(1710) & \rightarrow & |G\rangle  + \frac{1}{\sqrt{2r^2+ 1}}| \sqrt{2} r
N +  S\rangle  \\
f_0(1500) & \rightarrow & \epsilon |G\rangle + \frac{1}{\sqrt{2r^2+ 1}}
|N -\sqrt{2} r S\rangle   \\
f_0(1370) & \rightarrow & -|G\rangle + \frac{1}{\sqrt{2r^2 + 1}}| \sqrt{2}
r
N + S\rangle   \\
\end{array}
\label{eq:rinfmix}
\end{equation}

The pattern of
$N$ and $S$ phases in equations~(\ref{eq:1}) and
(\ref{eq:2}), namely two constructive and
one destructive, emerge so long as $r>0$.
The lattice results of ref.~\cite{Wein} imply $r = 1.20 \pm 0.07$. It is
for
these reasons,
 {\it inter alia}, that the output of refs.~\cite{re:AC,AC96,re:CFL,Wein}
and \cite{mix,ckmix}
are rather similar. In contrast, ref.~\cite{newshen}
finds opposite phases to the above and
this is because their mass matrix has $r<0$, which would
correspond to the mixing being driven by octet. This differs
radically from what would be expected for mixing driven by a glueball.
In the flavour symmetry limit, a glueball transforms as a flavour singlet;
there is a small octet component, if the above
results of lattice QCD~\cite{Wein} are a guide, but the idea that it should
be orthogonal to this and dominantly octet seems bizarre.
We note the mathematical consistency whereby if $r \to -r$ in
ref.~\cite{newshen}, their conclusions and results would parallel those
reported here, but for the reasons just outlined,  this
is so far from the lattice expectation that we do not discuss it
further.

The sharing of the glueball intensity among the three states
is driven by the proximity of the glueball to the bare states, amplified
by their $n\bar{n}$ contents (due to the factor $\sqrt{2}$ relative
amplitude for coupling to $n\bar{n}$ versus $s\bar{s}$). Apart from
this, the overall qualitative pattern of phases
 makes it hard to distinguish among
them. So the debate about whether the bare glueball lies
within~\cite{re:AC,AC96}
or above~\cite{Wein}
a prominent $q\bar{q}$ nonet may be academic unless fine measurement of the
quantitative rather than simply qualitative pattern can be extracted from
data.
However, this robust general feature of the phase pattern
enables this picture of glueball-nonet
mixing to be
disproved if their
common implications fail empirically.

In this context we draw attention to the non-trivial implications of
these dynamics for the flavour content of the $f_0(1500)$. While there
has been considerable debate about the nature of this state, there is
rather general agreement empirically
that the flavour content of the $f_0(1500)$ has
$N$ and $S$ out of phase. It is interesting that this emerges naturally,
and as a necessary consequence, for the ``middle" member when $G-q\bar{q}$
mixing is involved. While not a proof, this adds weight to the hypothesis
that
the $f_0(1500)$ is in a trio, with one partner higher and one lower in
mass.

Conversely, had the $f_0(1500)$ not had these characteristics
then this dynamics could have
been eliminated.

Since those mass mixings were first discussed,
there have emerged extensive data on the flavour-dependence of
these states' decays into pairs of pseudoscalar mesons. Analysis of
these decays can be used to give measures of the flavour composition
of these scalars, which bear no a priori relation to the mass mixing
arguments.
As such they provide an independent check on the above.
We shall now examine this
in section 3.

\section{Mixing and Decays}

The WA102 Collaboration at CERN has  published a complete set
of decay branching ratios for the $f_0(1370),f_0(1500)$ and $f_0(1710)$
to all kinematically allowed combinations of pairs of
pseudoscalars~\cite{etaetapap}.
These relative strengths depend upon the flavour content of the
scalars. The challenge is to decode this information and to compare the
resulting
pattern with that deduced from the mass mixings above.

We shall use the WA102 data in our primary analysis.
If instead we use a world average, our conclusions
are stable (shown in section 3.1.1).
In order to reduce model dependence, we shall take intuition
from the flux tube model~\cite{IP}, which is based on lattice QCD. This
suggests three major pathways for triggering the decays~\cite{AC96},

(i) the direct coupling of the quarkonia
component of the three states
to the final
pseudoscalar mesons ($PP$) (fig.~\ref{fi:1}a),

(ii) the decay of $gg \to qq \overline {qq}$
as in fig.~\ref{fi:1}b.
The resulting amplitudes can be obtained from eqs.~(A4) of ref.
{}~\cite{re:AC}
and have overall strength
$r_2$ (to be fitted) relative to the mode (i)~\cite{mix,ckmix}.

Finally, following ref.~\cite{re:AC,gershtein}, we allow for
 (iii) in fig.~\ref{fi:1}c,
 the direct coupling of the glue in the initial state to
isoscalar mesons (i.e. $\eta \eta$ and $\eta \eta^\prime$ decays)
and allow
$r_3$ to be the ratio
of mode (iii) to (i)~\cite{mix,ckmix}.

\par

In order to unfold the production
kinematics we use the invariant decay couplings~($\gamma_{ij}$)
for the observed decays, namely
we express the partial width~($\Gamma_{ij}$) as~\cite{re:AC}
\begin{equation}
\Gamma_{ij}=\gamma^2_{ij}|F_{ij}(\vec{q})|^2S_p(\vec{q})
\label{eq:a}
\end{equation}
where $S_p(\vec{q})$ denotes the phase space and $F_{ij}(\vec{q})$
are form factors appropriate to exclusive two body decays.
 Here we have followed ref.~\cite{re:AC}
and have chosen
\begin{equation}
|F_{ij}(\vec{q})|^2 = q^{2l}exp(-q^2/8\beta^2)
\label{eq:b}
\end{equation}
where $l$ is the angular momentum of the final state with daughter momenta
$q$
and we have used $\beta$~=~0.5~GeV/c~\cite{re:AC}.
The $f_0(1500)$ lies very near to threshold in the $\eta \eta^\prime$
decay mode, therefore we have used an average value of $q$ (190~MeV/c)
derived from a fit to the $\eta \eta^\prime$ mass spectrum.
\par
The branching ratios measured by the WA102 experiment for
the $f_0(1370)$, $f_0(1500)$ and $f_0(1710)$ are given in
table~\ref{ta:A1}.

\par
For quarkonium states the invariant couplings are dependent on the
flavour mixing angle $|Q\bar{Q}\rangle \equiv \cos \theta |N\rangle
- \sin \theta |S\rangle$.
Figs.~\ref{fi:2}, \ref{fi:3} and \ref{fi:4} show a plot of the
ratio of the invariant couplings as a function of $\theta$
for the $f_0(1370)$, $f_0(1500)$ and $f_0(1710)$ respectively.
%$\theta$ is related to the normal pseudoscalar mixing angle $\phi$
%by $\theta$~=~54.7$^\circ$+$\phi$.
Superimposed on the plot are the measured ratios with the $\pm 1 \sigma$
limits shown shaded.

As can be seen it is possible to find a solution for the $f_0(1370)$
and $f_0(1500)$ for small values of $\theta$
corresponding to them having a large $N \equiv n\bar{n}$ content. This is
already
an indication that they cannot both be members of the same $q\bar{q}$
nonet unless further degrees of freedom are present. It could be that the
$f_0(1370)$ belongs to a lower multiplet than the $f_0(1500)$
or that it does not exist~\cite{klempt}.
Even were either of these the case, there
would be need for a partner to the $f_0(1500)$ with $\theta \sim 90^\circ -
110^\circ$. Figure \ref{fi:4}
shows that the $f_0(1710)$ does not satisfy this.
 Thus we can already conclude the following:

(i) The $f_0(1500), f_0(1370)$
data show that if both of these states are real, they cannot be in the
same $q\bar{q}$ nonet without further degrees of freedom, such as a
glueball.

(ii) The $f_0(1710)$ data demonstrate
 the need to go beyond a simple $q \bar{q}$ picture at some point or that
data
 need to change.

Performing an elementary SU(3) calculation gives the
reduced partial widths in table~\ref{ta:A2},
where $\alpha=(\cos\phi-\sqrt{2}\sin\phi)/\sqrt{6}$,
$\beta=(\sin\phi+\sqrt{2}\cos\phi)/\sqrt{6}$, and
$\phi$ is the $q\bar{q}$ $S-N$ mixing angle of
$\eta$ and $\eta^\prime$. This mixing angle has been determined primarily
from
electromagnetic interactions that couple directly to the $q\bar{q}$
content of the $\eta, \eta^\prime$ states.
The relative importance of glue
coupling to $\eta$ and $\eta^\prime$ may be determined by gluon-driven
processes, such as $\psi \to \gamma \eta(\eta^\prime)$, or from theoretical
arguments about the coupling of the gluon
current to the $\eta^\prime$ system~\cite{novikov}. These independent
methods yield consistent results as follows.

(i) The ratio of $\psi$ radiative widths:

$$
\frac{\Gamma(\psi \to \eta^\prime \gamma)}{\Gamma(\psi \to \eta \gamma)}
 =    (\frac{q_{\eta^\prime}}{q_{\eta}})^3
|\frac{\langle 0|j|\eta^\prime \rangle}{ \langle 0|j|\eta\rangle }|^2
$$

 yields

$$
|\frac{g_{\eta^\prime}}{g_\eta}|
\equiv |\frac{\langle 0|j|\eta^\prime \rangle}{ \langle 0|j|\eta\rangle }|
= 2.5 \pm 0.2
$$

(ii) Theoretical arguments about the coupling of the gluon
current to the $\eta^\prime$ system~\cite{novikov} give

$$
 \frac{g_{\eta^\prime}}{g_\eta} = \frac{m_{\eta^\prime}^2}{\sqrt{2}
m_\eta^2} = 2.2
$$

We then perform a $\chi^2$ fit based on the branching ratios
given in table~\ref{ta:A1}, where we have required that the matrix $U$ in
eq.~(\ref{eq:e}) is
unitary, which applies an additional 6 constraints to the fit.
As input we use the masses of the $f_0(1500)$ and $f_0(1710)$.
In this way the nine parameters, $M_G$, $M_N$,
$M_S$, $M_3$, $f$, $r_2$, $r_3$, $r$ and $\phi$ are determined from the
fit. The mass of the $f_0(1370)$ is not well
established so we have left it as a free parameter ($M_3$).

\subsection{Flavour-Blind Glueball}

The parameters determined from the  fit are given in table~\ref{ta:A3}
and the fitted branching ratios together with the $\chi^2$
contributions of each are given in table~\ref{ta:A1}. Two robust
features merit immediate comment.
As can be seen the fit prefers a value of $r \approx$~1, in line with the
result of Lattice QCD~\cite{Wein}.
The mixing angles for the $\eta$, $\eta^\prime$ were unconstrained and
the fit chooses the canonical value of $-19^o$, in agreement with results
from
elsewhere.
The resulting flavour content of the mixed states is
\begin{equation}
\begin{array}{c c c c}
&f_{i1}^{(G)} & f_{i2}^{(S)} & f_{i3}^{(N)}\\
f_0(1710) & 0.39 & 0.91 & 0.14\\
f_0(1500) & -0.69 &  0.34 & -0.63\\
f_0(1370) & -0.62 &  0.13 & 0.77\\
\label{eq:j}
\end{array}
\end{equation}

This matrix confirms  the robustness of the qualitative pattern
that had followed from the mass matrix analyses, namely two states
with a singlet tendency and one with an octet. Although these relative
phases appear to be stable, the relative intensities of $G$ and flavours
differ;
the $m_G \sim 1440 \pm 10$ MeV is in the lower end of the mass range
preferred by
some reports from Lattice-QCD~\cite{re:lgt}, while rather lower than
the preferred solution of ref.~\cite{Wein}.
Consequently the leading structure of the mixing pattern follows from
degenerate perturbation theory with basis states $S, (N \pm G)/\sqrt{2}$.
The structure of eq. \ref{eq:j}, and what follows, all show this tendency.
 We shall discuss the implications of these results
in more detail later.
\par
This conclusion following from the decay analysis appears to be  stable
against changes in the detailed dynamics. As an example
 we return to the model assumption
made in our previous paper~\cite{ckmix},
namely that the glue couple to the $\eta$ states in proportion
to their $s\bar{s}$ content.
 The expressions for the partial widths
are given in
table~\ref{ta:B1}. (In this table we have corrected a sign
error that appeared in table 3 of
ref.~\cite{ckmix})).
The results of the fit using these expressions are given in
tables~\ref{ta:B2} and \ref{ta:B3}.
The
$\chi^2$ of the fit is 13.7
which is worse that the value obtained from the previous fit
but the structure essentially remains
unchanged:
\begin{equation}
\begin{array}{c c c c}
&f_{i1}^{(G)} & f_{i2}^{(S)} & f_{i3}^{(N)}\\
f_0(1710) & 0.42 & 0.89 & 0.16\\
f_0(1500) & -0.64 &  0.37 & -0.67\\
f_0(1370) & -0.63 &  0.15 & 0.76\\
\end{array}
\label{eq:j1}
\end{equation}

The general conclusion appears to be that analysis of decays of these
scalars reveals the same qualitative pattern of mixing phases as
deduced in the mass mixing analyses. The most general interpretation
is that these three states are mixed by a flavour singlet coupling: a
glueball
is a particular example of this. While this does not prove that the
glueball
is responsible, the robustness of the results, and the implications of
lattice
QCD that such a state should exist in this mass range, are strongly
suggestive.

\subsubsection{Insensitivity to choice of data sets}

To date we have used the branching ratios measured by experiment WA102
since this is the only experiment in the world to have measured all the
ratios.
However, the branching ratios have in part been measured by other
experiments.
Crystal Barrel~\cite{XBtot} have presented ratios for the
$f_0(1370)$ and $f_0(1500)$. BES have produced a measurement of the
$\pipi$/$KK$ ratio for the $f_0(1370)$~\cite{bugg1370} and there are
measurements of the
$\pipi$/$KK$ ratio for the $f_0(1710)$ from experiments WA76~\cite{WA76}
and  Mark III~\cite{dunwoodie}.
It is important to note that these other measured values
are compatible with the ones measured by experiment WA102.
\par
We have calculated the world average branching ratios for the $f_0(1370)$,
$f_0(1500)$ and the $f_0(1710)$ using all the available data. The values
are
given in table~\ref{ta:D1}. We have performed a fit to these values
using our formula and with $r$~=~1 and $\phi$~=~-19 degrees.
The parameters from the fit are given in table~\ref{ta:D2}.

The mixed states are
\begin{equation}
\begin{array}{c c c c}
&f_{i1}^{(G)} & f_{i2}^{(S)} & f_{i3}^{(N)}\\
f_0(1710) & 0.42 & 0.89 & 0.16\\
f_0(1500) & -0.66 &  0.37 & -0.64\\
f_0(1370) & -0.61 &  0.14 & 0.78\\
\end{array}
\label{eq:dataset}
\end{equation}
\noindent and the $\chi^2 = 14$. These are identical within the errors to
the results that followed from the fit to the WA102 data alone.

\subsubsection{Widths}

Anisovich \cite{anisovich} has argued that a signature of a glueball
driven mixing will be the presence of two states that are narrow and one
that
is broad. This result would arise if the flavour singlet channels that
drive

the mixing,
also dominate
the physical hadron states (in which case the ``octet" will be narrow due
to
decoupling, the
``glue + singlet" enhanced by constructive interference while the
``glue $-$ singlet" will be suppressed by destructive interference), but
it is less clear in a dynamics such as we have considered here. The
analyses of section 3 and 3.1
 do have implications for the relative sizes of the total widths of the
scalars for decays into pseudoscalar pairs. A consistency check on
these results should take account of this; that is the purpose of this
subsection.

The measured widths for the $f_0(1370)$, $f_0(1500)$ and $f_0(1710)$
are 272~$\pm$~40~$\pm$30~MeV,
108~$\pm$~14~$\pm$12~MeV and
124~$\pm$~16~$\pm$18~MeV respectively~\cite{etaetapap}.
Based on the observed decay modes of these states~\cite{etaetapap}
and taking into account the uncertainty in possible $\rho \rho$ modes,
which
would also imply the presence of $\omega \omega$ decay modes,
 the sums of the partial widths to pseudoscalar pairs are:

$$
f_0(1370) = 12~\pm~5~{\rm MeV };
f_0(1500) = 56~\pm~8~{\rm MeV };
f_0(1710) = 124~^{+16}_{-50}{\rm MeV}
$$

If in addition to the  analysis of section 3.1
we constrain the ratios of the observed total widths into pseudoscalar
pairs,
then we find an acceptable fit such that

$$
\frac{\Gamma(1710)}{\Gamma(1370)} = 7.1\pm2.2;
\frac{\Gamma(1500)}{\Gamma(1370)} =  10.0\pm3.0;
\frac{\Gamma(1710)}{\Gamma(1500)} = 0.7\pm0.2
$$

\noindent which are compatible with the empirical values above.
Performing a fit to the measured branching fractions gives the parameters
in table~\ref{ta:A1} and \ref{ta:A3}.

As can be seen,
adding the constraint of the ratio of the total widths
makes the $M_N$ and $M_G$ come very close together

$$
M_G = 1415 \;\;{\rm MeV} ;M_S = 1677 \;\;{\rm MeV} ;
M_N = 1402 \;\;{\rm MeV}
$$
\noindent  the mixed states are

% r= free
\begin{equation}
\begin{array}{c c c c}
&f_{i1}^{(G)} & f_{i2}^{(S)} & f_{i3}^{(N)}\\
f_0(1710) & 0.35 & 0.93 & 0.13\\
f_0(1500) & -0.61 &  0.29 & -0.74\\
f_0(1370) & -0.76 &  0.16 & 0.63\\
\label{eq:k}
\end{array}
\end{equation}
\noindent and the $\chi^2 = 10$.

There is an immediate physical reason for the pattern that emerges in
eq.~(\ref{eq:k}), namely the proximity of $m_G \sim m_N$.
In this case the parameters have the values $\xi > 1$ whereas
$\xi \omega \sim 1/3$. Thus mixing in the $G-N$ sector tends to be maximal
(analogous to eq.~(\ref{eq:infmix}) or eq.~(\ref{eq:rinfmix}))
 and the structure of the mixed states
will tend towards

\begin{equation}
\begin{array}{c c c c}
&f_{i1}^{(G)} & f_{i2}^{(S)} & f_{i3}^{(N)}\\
f_0(1710) & O(\xi\omega) & 1 & O(\sqrt{2}(\xi\omega)^2)\\
f_0(1500) & -\sqrt{1/3} & O(\xi\omega)  &  -\sqrt{2/3}\\
f_0(1370) & -\sqrt{2/3} &  O(\xi\omega) & \sqrt{1/3}\\
\label{eq:inflimit}
\end{array}
\end{equation}

 This structure is common to eq. (\ref{eq:k}) and indeed all of the mixing
patterns found throughout section 3.1 where the decay data constraints
have been imposed.
The central message of the decay data is that they prefer $m_G \sim m_N$.
%In view of
%this, we shall consider next the implications for other scenarios, where
$m_G$
%lies outside the mass ranged $m_S$ to $m_N$ spanned by the $q\bar{q}$
nonet
with
%which it is mixing.

\subsection{Flavour dependent $G \to q\bar{q}$}

\subsubsection{$m_G > m_S$ }

Our analysis of decays has pointed towards a $G-q\bar{q}$ coupling
that is approximately flavour independent, and a $m_G < \frac{m_S+m_N}{2}$.
This is in contrast to the
% lattice-QCD
analysis in ref.~\cite{Wein}
which
 preferred $m_S > m_G$.  In this section we ask what flavour
dependence of $G$ decays would be required for the latter solution
to emerge.

If we write
$R \equiv \frac{\gamma(G \to n\bar{n})}{\sqrt{2} \gamma(G \to s\bar{s})}$
then the reduced partial widths are given in table~\ref{ta:C1}.

$R \to 1$ recovers
the previous formulae, and initially we set $r_3 = 0$ (i.e. consider
only $G \to q\bar{q}$ and ignore any possible additional
direct coupling of $G \to \eta, \eta^\prime$).
Performing a fit to the measured branching fractions gives the parameters
in table~\ref{ta:C2} and \ref{ta:C3}.

The best fit (table~\ref{ta:C2}) has $r_2 \sim 5.4$
(which implies that the $G$ dominates
the decays) and $R \sim 0.67$ (which implies that $G$ couples more strongly
to $s\bar{s}$ than to $n\bar{n}$). This is what is required,
at least within the decay dynamics that we have assumed in this paper,
if the mass
matrix solution of ref.~\cite{Wein} is to be consistent with
the decay data. However, we note that the $\chi^2 = 80$. The major
mismatches
between fit and data are driven by $f_0(1710) \to \pi\pi/K\bar{K}$;
 $f_0(1710) \to \eta \eta^\prime/K\bar{K}$ and some from
$f_0(1500) \to \eta \eta^\prime/\eta\eta$. A challenge for future data will
be to
determine the accuracy of  these critical branching ratios.

We have investigated whether these conclusions are radically altered if
we allow $r_3$ to be free (i.e. allow additional  direct coupling of $G \to
\eta \eta^\prime$). These results also
are given in tables~\ref{ta:C2} and \ref{ta:C3}.
The $\chi^2$ falls to 19 and is significantly driven by the
$K\bar{K}/\pi\pi$ ratio being smaller (larger) than data for the
$f_0(1500)$ ($f_0(1370)$) respectively. $R \sim 0.5$ which still implies
a significant favouring of $G$ coupling to the heavier flavoured
$S$ rather than $N$.
With the advent of more powerful studies of QCD on the
lattice, it will be interesting to see if such behaviour is realised.
However, the $\chi^2$ is much larger than the value of 5.4
 that was found for the solution of table 1,
eq. (\ref{eq:j}) and section 3.1.

\subsubsection{Light Glueball: $m_G < m_N$}

In concluding our studies of flavour dependent $G$ couplings, we note that
if we allow the bare masses to be free and keep $r_3$~=~0,
then there exists a solution ($\chi^2 = 13 $)
with
$R = 1.4 \pm 0.4 $, for which
the mass of the bare glueball
is $m_G = 1310 \pm 14$MeV .
See tables~\ref{ta:C4} and \ref{ta:C5}.
The mixing matrix has the generic structure exhibited in
eq. (\ref{eq:inflimit}),
modulated by the $G,N$ tending to settle into the $f_0(1370)$ and
$f_0(1500)$
states. Explicitly it is

\begin{equation}
\begin{array}{c c c c}
&f_{i1}^{(G)} & f_{i2}^{(S)} & f_{i3}^{(N)}\\
f_0(1710) & 0.25 & 0.96 & 0.10\\
f_0(1500) & -0.37 &  0.13 & -0.92\\
f_0(1370) & -0.89 &  0.14 & 0.44\\
\end{array}
\label{eq:light}
\end{equation}

\noindent We do not discuss this further here, other than to note that
it implies that a light glueball
may be compatible with data. Furthermore, it is tantalising that such a
result could be in accord with Lattice QCD
(see for example the results with coarse
lattices in ref.~\cite{michael}).
If such a result should emerge from future studies of QCD with fine
grain lattices and including mixing  then a detailed analysis of the
phenomenology
along the lines we have instigated here would be most interesting.
We leave this as a future challenge for Lattice QCD.

If we then allow $r_3$ to be a free parameter we get a $\chi^2$ of 6.7.
The results are given in
tables~\ref{ta:C4} and \ref{ta:C5}. In this case $M_G$ is tending towards
$M_N$
and the solution is similar to the one we obtain in tables~\ref{ta:A1} and
\ref{ta:A3}.
The mixing matrix naturally shows the form of eq.~(\ref{eq:inflimit}).

%$$
%\begin{array}{c c c c}
%&f_{i1}^{(G)} & f_{i2}^{(S)} & f_{i3}^{(N)}\\
%f_0(1710) & 0.23 & 0.96 & 0.08\\
%f_0(1500) & -0.49 &  0.16 & -0.86\\
%f_0(1370) & -0.83 &  0.14 & 0.54\\
%\label{eq:l}
%\end{array}
%$$

%which again is seen to be similar to the one found in eq.~(\ref{eq:k})

We note that $m_G \sim 1402$ MeV is only slightly lower
than $m_N \sim 1446$ MeV
and the result is not dissimilar to that preferred in section 3.1.
However, in both
the cases $m_G > m_S$ and $m_G < m_N$ there is a clear omission, namely of
mixing
with nearest neighbour states above the glueball when $m_G > m_S$,
(section 3.2.1),
 or below it when
$m_N < m_G$, (section 3.2.2). Therefore, if $m_G$ should indeed
turn out to be $<m_N$, further analysis should be required involving the
$f_0(980)$ region, or the $\pi \pi$ $S-$wave continuum below 1
GeV~\cite{ochs}.

\section{Result}

Given the concerns expressed above about the $m_G > m_S$ and $m_G < m_N$
scenarios, it is the results of section 3.1 that are our preferred
solution.
With the hypotheses that the mixing describe the ratios of partial widths
for each individual resonance and also among the resonances, we take into
account the variability between WA102 data and world averages, and we allow
for
the uncertainties in flavour dependence of the glueball coupling. This
gives
our
final result, based on eqs. (\ref{eq:j}, \ref{eq:dataset}, \ref{eq:k}),
as follows.

\begin{equation}
\begin{array}{c c c c}
&f_{i1}^{(G)} & f_{i2}^{(S)} & f_{i3}^{(N)}\\
f_0(1710) & 0.39 \pm 0.03 & 0.91 \pm 0.02 & 0.15 \pm 0.02\\
f_0(1500) & -0.65 \pm 0.04 &  0.33 \pm 0.04 & -0.70 \pm 0.07\\
f_0(1370) & -0.69 \pm 0.07 &  0.15 \pm 0.01 & 0.70 \pm 0.07\\
\end{array}
\label{eq:result}
\end{equation}
\noindent and for which $m_G$ = $1443 \pm 24$ MeV, $m_N$ = $1377 \pm 20$
MeV

and $m_S$ = $1674 \pm 10$ MeV.

The specific numbers in the above matrix correlate with the specific values
of
$m_{G,N,S}$ but the generic structure shows the form of
eq.~(\ref{eq:inflimit}).
 Physically this reflects the dominant flavour-blind nature of
the $G-q\bar{q}$ coupling, amplified by the proximity of $m_G \sim m_N$
whereas
$m_G \neq m_S$. In the degenerate limit of $m_G \to m_N$, the mixing would
indeed
tend towards that in
eq.~(\ref{eq:inflimit}).

\section{Further Tests}

\subsection{$\gamma \gamma$ couplings}

The most sensitive probe of flavours and phases is potentially in $\gamma
\gamma$
couplings. The advantage is that
$\gamma \gamma$ couple to the $e^2$ of the flavours in amplitude and so the
net result is sensitive to the relative phases as well as the intensities.
That
this is a dominant dynamics is empirically well established for the $2^{++}
$
and $0^{-+}$ nonets; however it is moot whether it will in fact be so clean
for
the $0^{++}$. If it is dominant for this $J^{PC}$ also, then in the spirit
 of ref.~\cite{re:CFL}, ignoring mass-dependent
effects, the above imply

\begin{eqnarray}
\Gamma(f_0(1710)\rightarrow \gamma\gamma):\Gamma(f_0(1500)\rightarrow
\gamma\gamma):\Gamma(f_0(1370)\rightarrow \gamma\gamma)= \nonumber \\
(5z_1+\sqrt{2}y_1)^2:(5z_2+\sqrt{2}y_2)^2:(5z_3+\sqrt{2}y_3)^2
\label{eq:m}
\end{eqnarray}

For the case of the flavour blind glueball given in section 3.1
we get two predictions for these relationships:
one for the case when we do not add the total widths as a constraint and
one
when we do. We have averaged these two values and
used their difference as a measure of the systematic error
\begin{eqnarray}
\Gamma^{\gamma\gamma}(f_0(1710):f_0(1500):f_0(1370))
=
4.1 \pm 0.9\pm 0.3:9.7 \pm 0.9\pm 2.0:14.6 \pm 0.9 \pm 2.0
\end{eqnarray}

\par
The $\gamma \gamma$
width of $f_0(1500)$ exceeding that of $f_0(1710)$ arises
because the glueball is nearer to the $N$ than
the $S$. The pattern is radically different if nature chooses
 $G$ near to (or
even above) the
$S$, in which case the $f_0(1500)$ has the smallest $\gamma \gamma$
coupling of the three states \cite{re:CFL}. For example,
in the case of flavour dependent mixing with $M_G$ $>$ $M_S$ (section
3.2.1)
 we find
\begin{eqnarray}
\Gamma^{\gamma\gamma}(f_0(1710):f_0(1500):f_0(1370))
=
6.4 \pm 1.1:0.6  \pm 0.2:23.8 \pm 2.2
\end{eqnarray}
\noindent Contrast this with
 the case of flavour dependent mixing with $M_G$ $<$ $M_N$ (section 3.2.2)
for which
\begin{eqnarray}
\Gamma^{\gamma\gamma}(f_0(1710):f_0(1500):f_0(1370))
=
3.2 \pm 1.1:16.3  \pm 1.8:9.0 \pm 0.8
\end{eqnarray}

 This shows how these $\gamma \gamma$
couplings have the potential to pin down the input pattern. However, we
note
a caution with regards to $\gamma \gamma$ couplings necessarily being
the arbiter on $G-q\bar{q}$ wavefunctions in the $0^{++}$ partial waves.
 A problem here is that $0^{++}$ states decay to meson pairs in S-wave
(this is kinematically forbidden for the low-lying $0^{-+}$ or $2^{++}$
nonets)
 and so meson loops
may be expected to intercede between the $\gamma \gamma$ and $q\bar{q}$
levels.
Insights from $\gamma \gamma \to f_0(980)/a_0(980)$, from models and
ultimately
from lattice QCD will be needed to establish how clean in practice the
$\gamma \gamma$ measurements can be in the $0^{++}$ sector.

\subsection{Glue and Pomeron induced reactions: Central Production}

Our preferred  solutions
have two further implications for the production
of these states
in $p \bar{p}$ annihilations,
in central $pp$ collisions and in radiative $J/\psi$ decays
that are in accord with data. These are interesting in that they
are consequences of the output and were not used as constraints.

The production of the $f_0$ states in $p \overline p \to \pi + f_0$ is
expected
to be dominantly through the $N \equiv n \overline n$ components of the
$f_0$ state, possibly through $G$, but not prominently through
the $S \equiv s \overline s$ components. (The possible presence of hidden
$s \overline s$ at threshold, noted by \cite{ellis} is in general
swamped by the above, and in any event appears unimportant in flight).
The above mixing pattern implies that

\begin{equation}
\sigma(p \overline p \to \pi + f_0(1710)) <
\sigma(p \overline p \to \pi +  f_0(1370)) \sim
\sigma(p \overline p \to \pi +  f_0(1500))
\end{equation}
Experimentally~\cite{thoma} the relative production rates are,

\begin{equation}
p \overline p \to \pi + f_0(1370) : \pi + f_0(1500)) \sim
1 : 1.
\end{equation}
and there is no evidence for the production of the $f_0(1710)$.
This would be natural if the production were via the
$n \overline n$ component.
The actual magnitudes would however be model dependent; at this stage we
merely note the consistency of the data with the results of the mixing
analysis above.

For central production,
the cross sections of well established
quarkonia in WA102 suggest that the
production of $s \overline s$ is strongly suppressed \cite{pipikkpap}
relative to $n \overline n$.
The relative
cross sections for
the three states of interest here are

\begin{equation}
p p \to pp + ( f_0(1710):
 f_0(1500) :f_0(1370)) \sim 0.14:1.7:1.
\end{equation}
This would be natural if the production were via the
$N$ and $G$ components in phase.
\par
In addition,
the WA102 collaboration has studied the
production of these states as a function of
the azimuthal angle $\phi$, which is defined as the angle between the $p_T$
vectors of the two outgoing protons.
An important qualitative characteristic of these data is that
the $f_0(1710)$ and $f_0(1500)$ peak as
$\phi \to 0$ whereas the $f_0(1370)$ is more peaked
as $\phi \to 180$~\cite{WAphi}.
If the $G$ and $N$
components are produced coherently as $\phi \to 0$ but out of phase
as $\phi \to 180$, then  this pattern of $\phi$ dependence and relative
production rates would follow; however, the relative coherence of
$G$ and $N$ requires a dynamical explanation.
We do not have such an explanation and open this for debate.

\par
In $J/\psi$ radiative decays, the absolute rates depend
sensitively on the phases and relative strengths of the
$G$ relative to the $q \overline q$ component, as well as the
relative phase of $n\bar{n}$ and $s\bar{s}$ within the latter.
The general pattern though is clear. Following the discussion
in ref.~\cite{re:CFL} we expect that the coupling to $G$ will be large;
coupling to $q\bar{q}$ with ``octet tendency" will be suppressed; coupling
to $q\bar{q}$ with ``singlet tendency" will be intermediate. Hence the
rate for $f_0(1370)$ will be smallest as the $G$ interferes destructively
against the $q\bar{q}$ with ``singlet tendency". Conversely, the
$f_0(1710)$ is enhanced by their constructive interference. The $f_0(1500)$
contains $q\bar{q}$ with ``octet tendency" and its production will be
driven
dominantly by its $G$ content. If the $G$ mass is nearer to the $N$ than
to the $S$, as our results suggest, the $G$ component in $f_0(1500)$ is
large and causes the $J/\psi \to \gamma f_0(1500)$ rate to be comparable to
$J/\psi \to \gamma f_0(1710)$.
\par
In ref.~\cite{dunwoodie}, the branching ratio of
BR$(J/\psi \rightarrow \gamma f_0)(f_0\rightarrow \pi\pi + K \bar{K})$
for the $f_0(1500)$ and $f_0(1710)$ is presented. Using the WA102
measured
branching fractions~\cite{etaetapap}
for these resonances and assuming that all
major decay modes have been observed, the total relative production
rates in radiative $J/\psi$ decays can be calculated to be:
\begin{equation}
J/\psi \rightarrow f_0(1500) : J/\psi \rightarrow f_0(1710)
= 1.0 : 1.1 \pm 0.4
\end{equation}
which is consistent with the prediction above based on
our mixed state solution.
\par
In these mixed state solutions,
both
the $f_0(1500)$ and $f_0(1710)$ have
$N$ and $S$ contributions and so
it would be expected that both would be produced in
$\pi^-p$ and $K^-p$ interactions.
The $f_0(1500)$
has clearly been observed in
$\pi^-p$ interactions: it was first observed in the
$\eta\eta$ final state, although at that time it was referred to as
the $G(1590)$~\cite{NA12ETAETA}.
There is also evidence for the production
of the $f_0(1500)$ in $K^-p \rightarrow K^0_SK^0_S
\Lambda$~\cite{8GEV,LASS}.
The signal is much weaker compared to the
well known $s \bar{s}$ state the $f_2^\prime(1525)$, as expected with our
preferred mixings in eq. \ref{eq:result}
and the suppressed $K\bar{K}$ decay associated
with the
destructive $n\bar{n} - s\bar{s}$ phase in the wavefunction.
\par
There is evidence for the $f_0(1710)$ in the reaction
$\pi^-p \rightarrow K^0_SK^0_Sn$,
originally called the $S^{*\prime}(1720)$~\cite{ETKIN,BOLONKIN}.
One of the longstanding problems of the $f_0(1710)$ is that
in-spite of its dominant $K \bar{K}$ decay mode it was
not observed in $K^-p$ experiments~\cite{LASS,GAV}.
However, these concerns were based on the fact that initially the
$f_0(1710)$ had $J$~=~2.
In fact,
in ref.~\cite{lindenbaum} it was demonstrated that
if the $f_0(1710)$ had $J$~=~0, as it has now been found to have,
then the contribution in $\pi^-p$ and $K^-p$ are compatible.
One word of caution should be given here: the analysis
in ref.~\cite{lindenbaum} was performed with a $f_0(1400)$ rather than
the $f_0(1500)$ as we know it today. As a further test of
our solution, it would
be nice to see the analysis of ref.~\cite{lindenbaum}
repeated with the mass and width of the
$f_0(1500)$ and the decay parameters of the $f_0(1710)$
determined by the WA102 experiment.

\section{Conclusions}

We took as our guide the prediction of Lattice-QCD that, in the
quenched approximation, $m_G(0^{++}) \sim 1.5$ GeV,
and we explored the implications of
the hypothesis that this glueball mixes with its nearest $q\bar{q}$
neighbours.
This led us naturally to focus on the physical states the
$f_0(1370), f_0(1500)$
and $f_0(1710)$. This has been the philosophy behind several recent
analyses,
which appear different in detail at first sight, but which turn out
to have certain robust common features. We have abstracted these and
specified the critical data that are now required to make further
 progress.

The first studies of mixing were based on the mass matrix and the
assumption that the glueball-$q\bar{q}$ mixing is dominantly singlet in
character. The resulting output of two states that have constructive
interference
``singlet tendency" and one that has destructive ``octet tendency" is then
general. This can be seen as a common feature of
\cite{re:AC,Wein,mix}.

The absolute values of the flavour content are correlated with
the assumed masses of the bare glueball and quarkonium states.
Weingarten's initial work on the lattice assumed that the glueball
was higher in mass than the $s\bar{s}$ member of the
 nonet; this led to
a large glueball component in the heaviest state, the $f_0(1710)$ -
eq.~(5) and a large
$s\bar{s}$ content for the $f_0(1500)$.
Close and Amsler in contrast assumed that the glueball was initially
at a mass spanned by the nonet. This led to a different apportioning
of the glue among the states, eq.~(6),
in particular the $G$ and $s\bar{s}$ have
similar intensities in $f_0(1500)$ and $f_0(1710)$.

Subsequent work has also considered the decays into pseudoscalar pairs.
The qualitative features of the mixing are preserved, essentially due to
the assumed singlet dominance of the (glueball) mixing. A general feature
of these later works has been the assumption that the glueball component
of the wavefunctions has flavour independent couplings; any deviation
from this in the decays of the physical states is then due to the
glueball-flavour mixed eigenstates.

A common feature of the various solutions in eqs.~(5,6,\ref{eq:j}) and
(\ref{eq:k})
 is

(1) the $f_0(1370)$ has large $n\bar{n}$, small $s\bar{s}$
and significant $G$ content

(2) the $f_0(1710)$ has a large $s\bar{s}$ content in all
except Weingarten (eq.~(5)) whose solution instead has a large $G$

(3) $f_0(1500)$,as the central member of the trio,
  has $s\bar{s}$ and $n\bar{n}$ out of phase.

The decay analyses, eqs.~(\ref{eq:j}) and (\ref{eq:k})
 do show a systematic shift relative to the
original mass matrix analyses, eqs.~(5) and (6).
 This appears in two noticeable ways:

(1) The decay analyses want more $S$ in the $f_0(1700)$ and more
$G$ in the $f_0(1370)$. This is correlated to them wanting a rather
 light $G$ mass, whereby the $G$ mixes primarily with $N$, leaving the
``distant" $f_0(1700)$ as $S$ in leading order with a
10-20 \% $G$ intensity.

(2) A corollary is that
the $S$ content of the $f_0(1500)$ tends to be driven smaller
by the decay analyses. This is in marked contrast to Weingarten where
the $S$ content of the $f_0(1500)$ dominates, driven by the nearness of
$M_{s\bar{s}}$ to the physical eigenstate in his solution.

Therefore, if the $G$ decay is intrinsically flavour-blind, the results
of the decay analyses would imply that the $G$ is rather light,
nearer to the $N$ than to the $S$.
This is radically different to Weingarten's assumption that
$m_G>m_S>m_N$. The latter requires, within the assumptions of our
analysis, that $G$ couples to $S$ more strongly than to $N$, and also that
the coupling of $G \to $ meson pairs is stronger than $Q\bar{Q}$ to
the same meson pairs. This latter result appears unnatural to us. It will
be
a challenge to lattice QCD to study these couplings to see if there is any
sign of such unexpected behaviour. In the absence of such an anomaly, we
anticipate that the likely inference of this analysis is that
$G$ is rather light, nearer to $N$ than to $S$.

With the hypotheses that the mixing describe the ratios of partial widths
for each individual resonance and also among the resonances, allowing
for variability between WA102 data and world averages, and allowing for
the uncertainties in flavour dependence of the glueball coupling, the
results
of section 3 lead us to the following summary for the favoured result:

\begin{equation}
\begin{array}{c c c c}
&f_{i1}^{(G)} & f_{i2}^{(S)} & f_{i3}^{(N)}\\
f_0(1710) & 0.39 \pm 0.03 & 0.91 \pm 0.02 & 0.15 \pm 0.02\\
f_0(1500) & -0.65 \pm 0.04 &  0.33 \pm 0.04 & -0.70 \pm 0.07\\
f_0(1370) & -0.69 \pm 0.07 &  0.15 \pm 0.01 & 0.70 \pm 0.07\\
\end{array}
\label{final2}
\end{equation}

\noindent for which $m_G=1443 \pm 24$ MeV,
$m_N = 1377 \pm 20$ MeV and $m_S = 1674 \pm 10$ MeV.

We make two further comments about this result.

(i) In the
quenched approximation one would expect an $a_0$ state that
is mass
degenerate with the $N$ state before any mixing. Hence we would expect
the $a_0$ to be in this region of $1350 - 1400$ MeV.
The existence and mass of any $a_0$ other than the $a_0(980)$
is still controversial and
we advertise this as an important datum that could further constrain
analyses such as those we have made in this paper. The presence or absence
of
an $a_0$ in the mass region favoured by us could have implications
for the interpretation of the $a_0(980)$ and $f_0(980)$ states.
Establishing
the status of $a_0(\sim 1400)$ should be a high priority in the quest to
understand the nature of the $0^{++}$ mesons.

(ii) We also note that our result that $m_S - m_N \sim 300$ MeV is
consistent with what one
would expect from
$f_2(1525) - f_2(1270)$ or, equivalently, the naive
accounting of masses for constituent quarks where $2m_s - 2m_n \sim 0.3$
GeV.

\par
In summary,
based on the hypothesis that
the scalar glueball
mixes with the nearby $q \overline q$ nonet states,
we have determined the flavour content of the
$f_0(1370), f_0(1500)$ and $f_0(1710)$
by studying their decays into all pseudoscalar meson pairs.
It suggests that the $m_G$ is relatively light, nearer in
mass to $m_N$ than $m_S$.
The solution we have found
is also compatible with the relative production
strengths of the $f_0(1370), f_0(1500)$ and $f_0(1710)$
in $pp$ central production, $p \bar{p}$ annihilations
and $J/\psi$ radiative decays.

\begin{center}
{\bf Acknowledgements}
\end{center}
\par
%We are indebted to H.J. Lipkin for comments on heavy flavour decays.
We are indebted to C. Michael and M. Teper for discussions on glueballs
and mixing in lattice QCD.
This work is supported, in part, by grants from
the British Particle Physics and Astronomy Research Council,
the British Royal Society,
and the European Community Human Mobility Program Eurodafne,
contract NCT98-0169.

\newpage
\begin{table}[h]
\caption{The solutions
for the minimum $\chi^2$ (total width constraint).}
\label{ta:A1}
\vspace{0.2in}
\begin{center}
\begin{tabular}{|c|c|cc|cc|} \hline
 & & & && \\
 & Measured
 & \multicolumn{2}{c|}{new formula $r$ free}
 & \multicolumn{2}{c|}{new formula $r$ free width cons}\\
 & Branching &Fitted& $\chi^2$
 &Fitted& $\chi^2$\\
 & ratio& & && \\
 & & & && \\ \hline
  & & & && \\
 $\frac{f_0(1370)\rightarrow \pi \pi}{f_0(1370)\rightarrow K \overline K}$
&2.17 $\pm$ 0.9 & 2.14 & 0.001 &0.38 & 3.97 \\
 & & & && \\
 $\frac{f_0(1370)\rightarrow \eta \eta}{f_0(1370)\rightarrow K \overline K}
$
&0.35 $\pm$ 0.21 & 0.41 & 0.08  &0.42 & 0.13 \\
 & & & && \\
 $\frac{f_0(1500)\rightarrow \pi \pi}{f_0(1500)\rightarrow \eta \eta}$
&5.5 $\pm$ 0.84 & 5.79 & 0.12  &5.7 & 0.06\\
 & & & && \\
 $\frac{f_0(1500)\rightarrow K \overline K}{f_0(1500)\rightarrow \pi \pi}$
&0.32 $\pm$ 0.07 & 0.38 & 0.65  & 0.43 & 2.5\\
 & & & && \\
 $\frac{f_0(1500)\rightarrow \eta \eta^\prime}{f_0(1500)\rightarrow \eta
\eta}$
&0.52 $\pm$ 0.16 & 0.50 & 0.02 & 0.55 & 0.02 \\
 & & & && \\
 $\frac{f_0(1710)\rightarrow \pi \pi}{f_0(1710)\rightarrow K \overline K}$
&0.20 $\pm$ 0.03 & 0.18 & 0.43  &0.19 & 0.10\\
 & & & && \\
 $\frac{f_0(1710)\rightarrow \eta \eta}{f_0(1710)\rightarrow K \overline K}
$
&0.48 $\pm$ 0.14 & 0.20 & 4.08 &0.24 & 2.90\\
 & & & && \\
 $\frac{f_0(1710)\rightarrow \eta \eta^\prime}{f_0(1710)\rightarrow \eta
\eta}$
&$< 0.05 (90 \% cl)$  & 0.04 & 0.06  &0.03 & 0.05\\
 & & & && \\ \hline
\end{tabular}
\end{center}
\end{table}
\begin{table}[h]
\caption{The theoretical reduced partial widths (new formula).}
\label{ta:A2}
\vspace{0.2in}
\begin{center}
\begin{tabular}{|c|c|} \hline
 &  \\
$\gamma^2(f_i\rightarrow \eta\eta^\prime)$
&$2[2\alpha\beta(z_i - \sqrt{2}y_i)
+ 2\frac{g_{\eta^\prime}}{g_{\eta}} x_i r_3]^2$ \\
 &  \\
$\gamma^2(f_i\rightarrow \eta\eta)$
&$[2\alpha^2z_i+2\sqrt{2}\beta^2y_i+r_2x_i+2 x_i r_3]^2$ \\
 &  \\
$\gamma^2(f_i \rightarrow \pi\pi)$
&$3[z_i+r_2x_i]^2$ \\
 &  \\
$\gamma^2(f_i\rightarrow K\bar{K})$
&$4[\frac{1}{2}(z_i +\sqrt{2}y_i)+r_2x_i]^2$ \\
 & \\ \hline
\end{tabular}
\end{center}
\end{table}
\begin{table}[h]
\caption{The solutions
for the minimum $\chi^2$ new formula.}
\label{ta:A3}
\vspace{0.2in}
\begin{center}
\begin{tabular}{|c|c|c|}\hline
& & \\
Parameters& r free & r free width cons\\
& &  \\ \hline
& &  \\
$\chi^2$  & 5.4                & 10.1 \\
& &  \\
$M_G$ (MeV) &1441 $\pm$ 12     &1415 $\pm$ 16 \\
$M_S$ (MeV) &1675 $\pm$ 9      &1680 $\pm$ 12\\
$M_N$ (MeV) &1364 $\pm$ 19     &1405 $\pm$ 22\\
$M_3$ (MeV) &1264 $\pm$ 14     &1265 $\pm$ 18\\
$f$ (MeV)   &85 $\pm$ 10       &85 $\pm$ 12\\
$\phi$ (Deg)&-19 $\pm$ 3       &-15 $\pm$ 5\\
$r_2$       &0.96 $\pm$ 0.26   &1.21 $\pm$ 0.29\\
$r_3$       &0.09$\pm$0.03     &0.15$\pm$0.04  \\
$r$         & 1.0$\pm$ 0.3     & 0.96$\pm$ 0.3  \\
& & \\ \hline
\end{tabular}
\end{center}
\end{table}
\begin{table}[h]
\caption{The theoretical reduced partial widths (old formula corrected).}
\label{ta:B1}
\vspace{0.2in}
\begin{center}
\begin{tabular}{|c|c|} \hline
 &  \\
$\gamma^2(f_i\rightarrow \eta\eta^\prime)$
&$2[2\alpha\beta(z_i - \sqrt{2}y_i) - 2\alpha \beta x_i r_3]^2$ \\
 &  \\
$\gamma^2(f_i\rightarrow \eta\eta)$
&$[2\alpha^2z_i+2\sqrt{2}\beta^2y_i+r_2x_i+2\beta^2 x_i r_3]^2$ \\
 &  \\
$\gamma^2(f_i \rightarrow \pi\pi)$
&$3[z_i+r_2x_i]^2$ \\
 &  \\
$\gamma^2(f_i\rightarrow K\bar{K})$
&$4[\frac{1}{2}(z_i +\sqrt{2}y_i)+r_2x_i]^2$ \\
 & \\ \hline
\end{tabular}
\end{center}
\end{table}
\clearpage
\begin{table}[h]
\caption{The solutions
for the minimum $\chi^2$ (old formula).}
\label{ta:B2}
\vspace{0.2in}
\begin{center}
\begin{tabular}{|c|c|}\hline
&   \\
Parameters & corrected \\
&   \\ \hline
&   \\
$\chi^2$  & 13.7  \\
&  \\
$M_G$ (MeV) &1438 $\pm$ 12  \\
$M_S$ (MeV) &1667 $\pm$ 10\\
$M_N$ (MeV) &1370 $\pm$ 19\\
$M_3$ (MeV) &1258 $\pm$ 28\\
$f$ (MeV) &95 $\pm$ 26\\
$\phi$ (Deg &-19$\pm$ 2\\
$r_2$ &0.94 $\pm$ 0.09\\
$r_3$ &0.40$\pm$ 0.30  \\
& \\ \hline
\end{tabular}
\end{center}
\end{table}
\newpage
\begin{table}[h]
\caption{The measured and predicted branching ratios
with the individual $\chi^2$ contributions coming from the fits.}
\label{ta:B3}
\vspace{0.2in}
\begin{center}
\begin{tabular}{|c|c|cc|} \hline
 & & & \\
 & Measured
 & \multicolumn{2}{c|}{Old formula corrected} \\
 & Branching &Fitted& $\chi^2$  \\
 & ratio& & \\
 & & & \\ \hline
  & & & \\
 $\frac{f_0(1370)\rightarrow \pi \pi}{f_0(1370)\rightarrow K \overline K}$
&2.17 $\pm$ 0.9 & 2.1 & 0.005 \\
 & & & \\
 $\frac{f_0(1370)\rightarrow \eta \eta}{f_0(1370)\rightarrow K \overline K}
$
&0.35 $\pm$ 0.21 & 0.18 & 0.62 \\
 & & & \\
 $\frac{f_0(1500)\rightarrow \pi \pi}{f_0(1500)\rightarrow \eta \eta}$
&5.5 $\pm$ 0.84 & 6.5 & 1.4 \\
 & & & \\
 $\frac{f_0(1500)\rightarrow K \overline K}{f_0(1500)\rightarrow \pi \pi}$
&0.32 $\pm$ 0.07 & 0.32 & 1.1 \\
 & & & \\
 $\frac{f_0(1500)\rightarrow \eta \eta^\prime}{f_0(1500)\rightarrow \eta
\eta}$
&0.52 $\pm$ 0.16 & 0.17 & 4.8 \\
 & & & \\
 $\frac{f_0(1710)\rightarrow \pi \pi}{f_0(1710)\rightarrow K \overline K}$
&0.20 $\pm$ 0.03 & 0.2 & 0.03 \\
 & & & \\
 $\frac{f_0(1710)\rightarrow \eta \eta}{f_0(1710)\rightarrow K \overline K}
$
&0.48 $\pm$ 0.14 & 0.19 & 4.3 \\
 & & & \\
 $\frac{f_0(1710)\rightarrow \eta \eta^\prime}{f_0(1710)\rightarrow \eta
\eta}$
&$< 0.05 (90 \% cl)$& 0.09 & 2.5   \\
 & & & \\ \hline
\end{tabular}
\end{center}
\end{table}
\newpage
\begin{table}[h]
\caption{Our formula (no total width) world average.}
\label{ta:D1}
\vspace{0.2in}
\begin{center}
\begin{tabular}{|c|c|cc|} \hline
 & & &  \\
 & Measured
 & \multicolumn{2}{c|}{all} \\
 & Branching &Fitted& $\chi^2$ \\
 & ratio& &  \\
 & & &   \\ \hline
  & & &   \\
 $\frac{f_0(1370)\rightarrow \pi \pi}{f_0(1370)\rightarrow K \overline K}$
&1.78 $\pm$ 0.9 & 2.16 & 0.18 \\
 & & &   \\
 $\frac{f_0(1370)\rightarrow \eta \eta}{f_0(1370)\rightarrow K \overline K}
$
&0.11 $\pm$ 0.15 & 0.27 & 1.1 \\
 & & &  \\
 $\frac{f_0(1500)\rightarrow \pi \pi}{f_0(1500)\rightarrow \eta \eta}$
&7.7 $\pm$ 1.5 & 8.3 & 0.16 \\
 & & &  \\
 $\frac{f_0(1500)\rightarrow K \overline K}{f_0(1500)\rightarrow \pi \pi}$
&0.21 $\pm$ 0.05 & 0.35 & 7.4 \\
 & & &  \\
 $\frac{f_0(1500)\rightarrow \eta \eta^\prime}{f_0(1500)\rightarrow \eta
\eta}$
&0.71 $\pm$ 0.13 & 0.0.56 & 1.4 \\
 & & &  \\
 $\frac{f_0(1710)\rightarrow \pi \pi}{f_0(1710)\rightarrow K \overline K}$
&0.26 $\pm$ 0.07 & 0.21 & 0.44 \\
 & & &  \\
 $\frac{f_0(1710)\rightarrow \eta \eta}{f_0(1710)\rightarrow K \overline K}
$
&0.48 $\pm$ 0.14 & 0.0.22 &3.5 \\
 & & & \\
 $\frac{f_0(1710)\rightarrow \eta \eta^\prime}{f_0(1710)\rightarrow \eta
\eta}$
&$< 0.05 (90 \% cl)$  & 0.05 & 0.01 \\
 & & & \\ \hline
\end{tabular}
\end{center}
\end{table}
\begin{table}[h]
\caption{Our formula (no total width) world average.}
\label{ta:D2}
\vspace{0.2in}
\begin{center}
\begin{tabular}{|c|c|}\hline
&   \\
Parameters& all ratios \\
&   \\ \hline
&   \\
$\chi^2$  &14.2 \\
& \\
$M_G$ (MeV) &1473 $\pm$ 15 \\
$M_S$ (MeV)&1667 $\pm$ 16 \\
$M_N$ (MeV)&1363 $\pm$ 21 \\
$M_3$ (MeV)&1258 $\pm$ 33 \\
$f$ (MeV)&94 $\pm$ 16\\
$\phi$ (Deg)&-19  \\
$r_2$ & 0.99 $\pm$ 0.27\\
$r_3$ & 0.04 $\pm$ 0.02\\
& \\ \hline
\end{tabular}
\end{center}
\end{table}
\begin{table}[h]
\caption{The theoretical reduced partial widths (Weingarten).}
\label{ta:C1}
\vspace{0.2in}
\begin{center}
\begin{tabular}{|c|c|} \hline
 &  \\
$\gamma^2(f_i\rightarrow \eta\eta^\prime)$
&$2[2\alpha\beta(z_i - \sqrt{2}y_i)
+ x_i r_2(1-R^{-2})+ 2\frac{g_{\eta^\prime}}{g_{\eta}} x_i r_3]^2$ \\
 &  \\
$\gamma^2(f_i\rightarrow \eta\eta)$
&$[2\alpha^2(z_i+x_ir_2)+ 2\beta^2(\sqrt{2}y_i+r_2x_iR^{-2})
+2x_i r_3]^2$ \\
 &  \\
$\gamma^2(f_i \rightarrow \pi\pi)$
&$3[z_i+r_2x_i]^2$ \\
 &  \\
$\gamma^2(f_i\rightarrow K\bar{K})$
&$4[\frac{1}{2}(z_i +\sqrt{2}y_i)+R^{-1} r_2x_i]^2$ \\
 & \\ \hline
\end{tabular}
\end{center}
\end{table}
\begin{table}[h]
\caption{The solutions
for the minimum $\chi^2$ (weingartens formula fixed mass $R$ $<$ 1 ).}
\label{ta:C2}
\vspace{0.2in}
\begin{center}
\begin{tabular}{|c|c|c|}\hline
& &  \\
Parameters& $r_3$~=~0 & $r_3$ free \\
& &  \\ \hline
& &  \\
$\chi^2$  &81.7 & 19.3  \\
& &  \\
$M_G$ (MeV) &1622 &1622  \\
$M_S$ (MeV)&1514  &1514\\
$M_N$ (MeV)&1470  &1470\\
$M_3$ (MeV)&1366  &1363\\
$f$ (MeV)&88 $\pm$ 21&99 $\pm$ 21\\
$\phi$ (Deg)&-19  &-19\\
$r_2$ & 5.40 $\pm$ 0.94&2.89 $\pm$ 0.40\\
$r_3$ & 0.0  &1.08$\pm$ 0.17  \\
$R$ & 0.68 $\pm$ 0.16&0.52$\pm$ 0.04  \\
& &  \\ \hline
\end{tabular}
\end{center}
\end{table}
\newpage
\begin{table}[h]
\caption{Weingarten formula fixed mass $R$ $<$ 1   }
\label{ta:C3}
\vspace{0.2in}
\begin{center}
\begin{tabular}{|c|c|cc|cc|} \hline
 & & & && \\
 & Measured
 & \multicolumn{2}{c|}{$r_3$ fixed 0}
 & \multicolumn{2}{c|}{$r_3$ free }\\
 & Branching &Fitted& $\chi^2$
 &Fitted& $\chi^2$\\
 & ratio& & && \\
 & & & && \\ \hline
  & & & && \\
 $\frac{f_0(1370)\rightarrow \pi \pi}{f_0(1370)\rightarrow K \overline K}$
&2.17 $\pm$ 0.9 & 0.34 & 4.1 &0.11 & 5.2  \\
 & & & && \\
 $\frac{f_0(1370)\rightarrow \eta \eta}{f_0(1370)\rightarrow K \overline K}
$
&0.35 $\pm$ 0.21 & 0.19 & 0.58  &0.49 & 0.43 \\
 & & & && \\
 $\frac{f_0(1500)\rightarrow \pi \pi}{f_0(1500)\rightarrow \eta \eta}$
&5.5 $\pm$ 0.84 & 6.5 & 1.41  &4.0 & 3.1 \\
 & & & && \\
 $\frac{f_0(1500)\rightarrow K \overline K}{f_0(1500)\rightarrow \pi \pi}$
&0.32 $\pm$ 0.07 & 0.39 & 1.00  & 0.16 & 5.3\\
 & & & && \\
 $\frac{f_0(1500)\rightarrow \eta \eta^\prime}{f_0(1500)\rightarrow \eta
\eta}$
&0.52 $\pm$ 0.16 & 0.0003 & 10.5 & 0.32 & 1.5 \\
 & & & && \\
 $\frac{f_0(1710)\rightarrow \pi \pi}{f_0(1710)\rightarrow K \overline K}$
&0.20 $\pm$ 0.03 & 0.39 & 39.5  &0.26 & 3.6\\
 & & & && \\
 $\frac{f_0(1710)\rightarrow \eta \eta}{f_0(1710)\rightarrow K \overline K}
$
&0.48 $\pm$ 0.14 & 0.22 & 3.48 &0.44 & 0.09\\
 & & & && \\
 $\frac{f_0(1710)\rightarrow \eta \eta^\prime}{f_0(1710)\rightarrow \eta
\eta}$
&$< 0.05 (90 \% cl)$  & 0.18 & 21.1  &0.05 & 0.003\\
 & & & && \\ \hline
\end{tabular}
\end{center}
\end{table}
\begin{table}[h]
\caption{The solutions
for the minimum $\chi^2$ (free mass $R$ $>$ 1 ).}
\label{ta:C4}
\vspace{0.2in}
\begin{center}
\begin{tabular}{|c|c|c|}\hline
& &  \\
Parameters& $r_3$ Fixed 0 & $r_3$ free \\
& &  \\ \hline
& &  \\
$\chi^2$  &12.9 & 6.7  \\
& &  \\
$M_G$ (MeV)  &1310 $\pm$ 14  &1402 $\pm$ 12  \\
$M_S$ (MeV)  &1692 $\pm$ 16  &1694 $\pm$ 13\\
$M_N$ (MeV)  &1460 $\pm$ 23  &1446 $\pm$ 18\\
$M_3$ (MeV)  &1257 $\pm$ 25  &1301 $\pm$ 23\\
$f$ (MeV)    &70 $\pm$ 11    &66 $\pm$ 10\\
$\phi$ (Deg) &-19  &-19\\
$r_2$        &1.69 $\pm$ 0.21&1.91 $\pm$ 0.20\\
$r_3$        &0.              &0.12$\pm$ 0.04  \\
$R$          &1.37$\pm$ 0.38  &1.32$\pm$ 0.32  \\
& &  \\ \hline
\end{tabular}
\end{center}
\end{table}
\newpage
\begin{table}[h]
\caption{free mass $R$ $>$ 1   }
\label{ta:C5}
\vspace{0.2in}
\begin{center}
\begin{tabular}{|c|c|cc|cc|} \hline
 & & & && \\
 & Measured
 & \multicolumn{2}{c|}{$r_3$ fixed 0}
 & \multicolumn{2}{c|}{$r_3$ free}\\
 & Branching &Fitted& $\chi^2$
 &Fitted& $\chi^2$\\
 & ratio& & && \\
 & & & && \\ \hline
  & & & && \\
 $\frac{f_0(1370)\rightarrow \pi \pi}{f_0(1370)\rightarrow K \overline K}$
&2.17 $\pm$ 0.9 &1.72 & 0.25   &1.57 & 0.44 \\
 & & & && \\
 $\frac{f_0(1370)\rightarrow \eta \eta}{f_0(1370)\rightarrow K \overline K}
$
&0.35 $\pm$ 0.21  &0.30 & 0.05   &0.38 & 0.05 \\
 & & & && \\
 $\frac{f_0(1500)\rightarrow \pi \pi}{f_0(1500)\rightarrow \eta \eta}$
&5.5 $\pm$ 0.84   &7.8 & 7.7   &6.2 & 0.6 \\
 & & & && \\
 $\frac{f_0(1500)\rightarrow K \overline K}{f_0(1500)\rightarrow \pi \pi}$
&0.32 $\pm$ 0.07  & 0.32 & 0.005  & 0.37 & 0.53\\
 & & & && \\
 $\frac{f_0(1500)\rightarrow \eta \eta^\prime}{f_0(1500)\rightarrow \eta
\eta}$
&0.52 $\pm$ 0.16  & 0.48 & 0.06   & 0.60 & 0.23 \\
 & & & && \\
 $\frac{f_0(1710)\rightarrow \pi \pi}{f_0(1710)\rightarrow K \overline K}$
&0.20 $\pm$ 0.03  &0.20 & 0.0003  &0.20 & 0.0003\\
 & & & && \\
 $\frac{f_0(1710)\rightarrow \eta \eta}{f_0(1710)\rightarrow K \overline K}
$
&0.48 $\pm$ 0.14  &0.17 & 4.75 &0.20 & 4.14\\
 & & & && \\
 $\frac{f_0(1710)\rightarrow \eta \eta^\prime}{f_0(1710)\rightarrow \eta
\eta}$
&$< 0.05 (90 \% cl)$  &0.05 & 0.025  &0.05 & 0.004\\
 & & & && \\ \hline
\end{tabular}
\end{center}
\end{table}
%%
%%%%%%%%%%%%%%%%%%%%%%%%%%%%%%%%%%%%%%%%%%%%%%%%%%%%%%%%%%%%%%%%%%%%%%%%%%

\clearpage
{ \large \bf Figures \rm}
\begin{figure}[h]
\caption{
The Decays to Pseudoscalar meson pairs ($PP$) considered in this analysis.
a) The coupling of the $q \bar{q}$
to the $PP$ pair,
b) the coupling of the glueball
component to $PP$ and c) the direct coupling of gluons to
the gluonic component of the final state mesons.
}
\label{fi:1}
\end{figure}
\begin{figure}[h]
\caption{
The ratio of the invariant coupling amplitudes squared as a function
of the flavour mixing angle $\theta$ for the $f_0(1370)$.
Superimposed on the plots is the measured ratios. The band indicates
the $\pm 1 \sigma$ region.
}
\label{fi:2}
\end{figure}
\begin{figure}[h]
\caption{The ratio of the invariant coupling amplitudes squared as a
function
of the flavour mixing angle $\theta$ for the $f_0(1500)$.
Superimposed on the plots is the measured ratios. The band indicates
the $\pm 1 \sigma$ region.
}
\label{fi:3}
\end{figure}
\begin{figure}[h]
\caption{
The ratio of the invariant coupling amplitudes squared as a function
of the flavour mixing angle $\theta$ for the $f_0(1710)$.
Superimposed on the plots is the measured ratios. The band indicates
the $\pm 1 \sigma$ region.
}
\label{fi:4}
\end{figure}
%%%%%%%%%

%%%%%%%%%
\begin{center}
\epsfig{figure=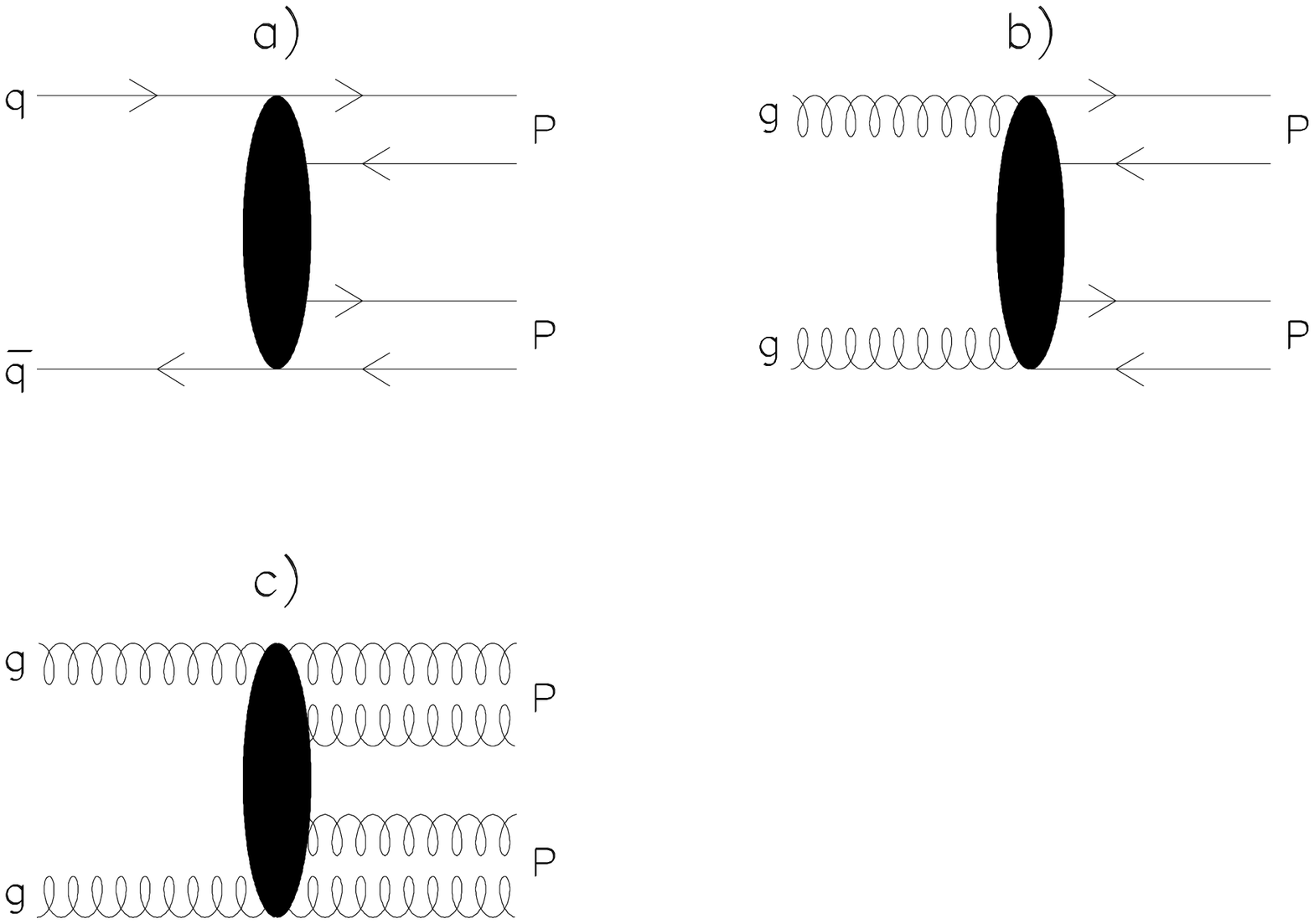,height=22cm,width=17cm}
\end{center}
\begin{center} {Figure 1} \end{center}
\newpage
\begin{center}
\epsfig{figure=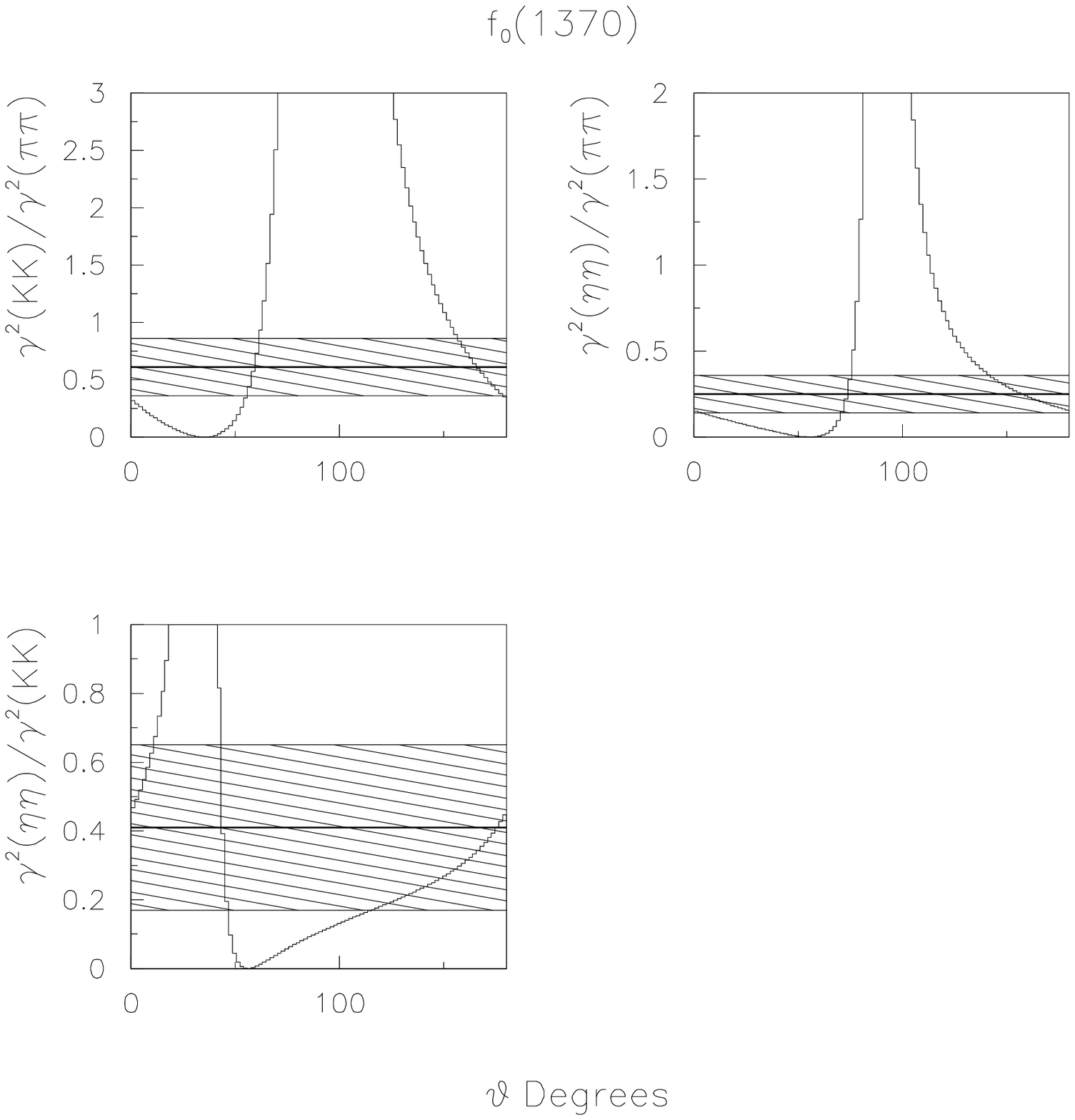,height=22cm,width=17cm}
\end{center}
\begin{center} {Figure 2} \end{center}
\newpage
\begin{center}
\epsfig{figure=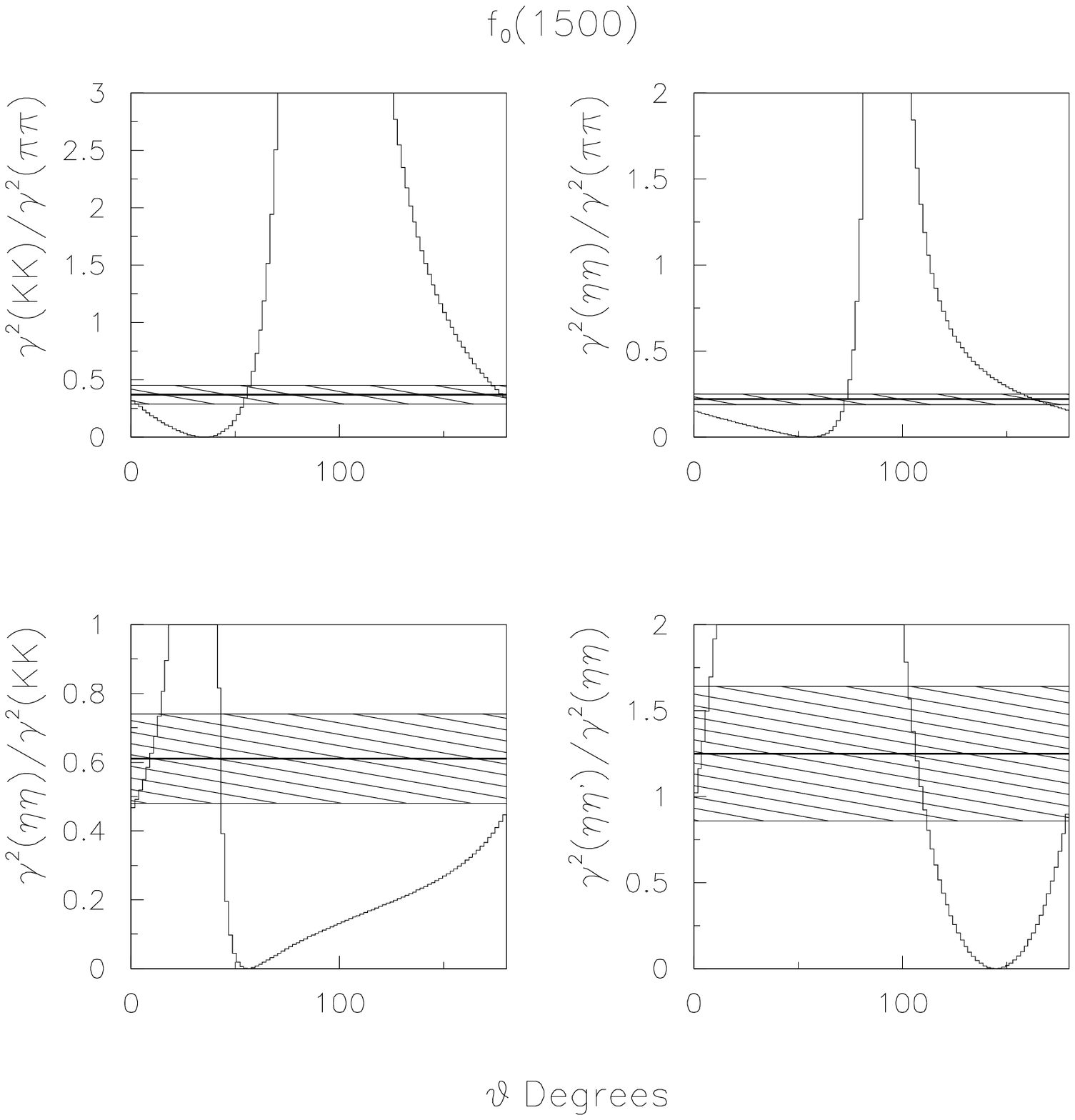,height=22cm,width=17cm}
\end{center}
\begin{center} {Figure 3} \end{center}
\newpage
\begin{center}
\epsfig{figure=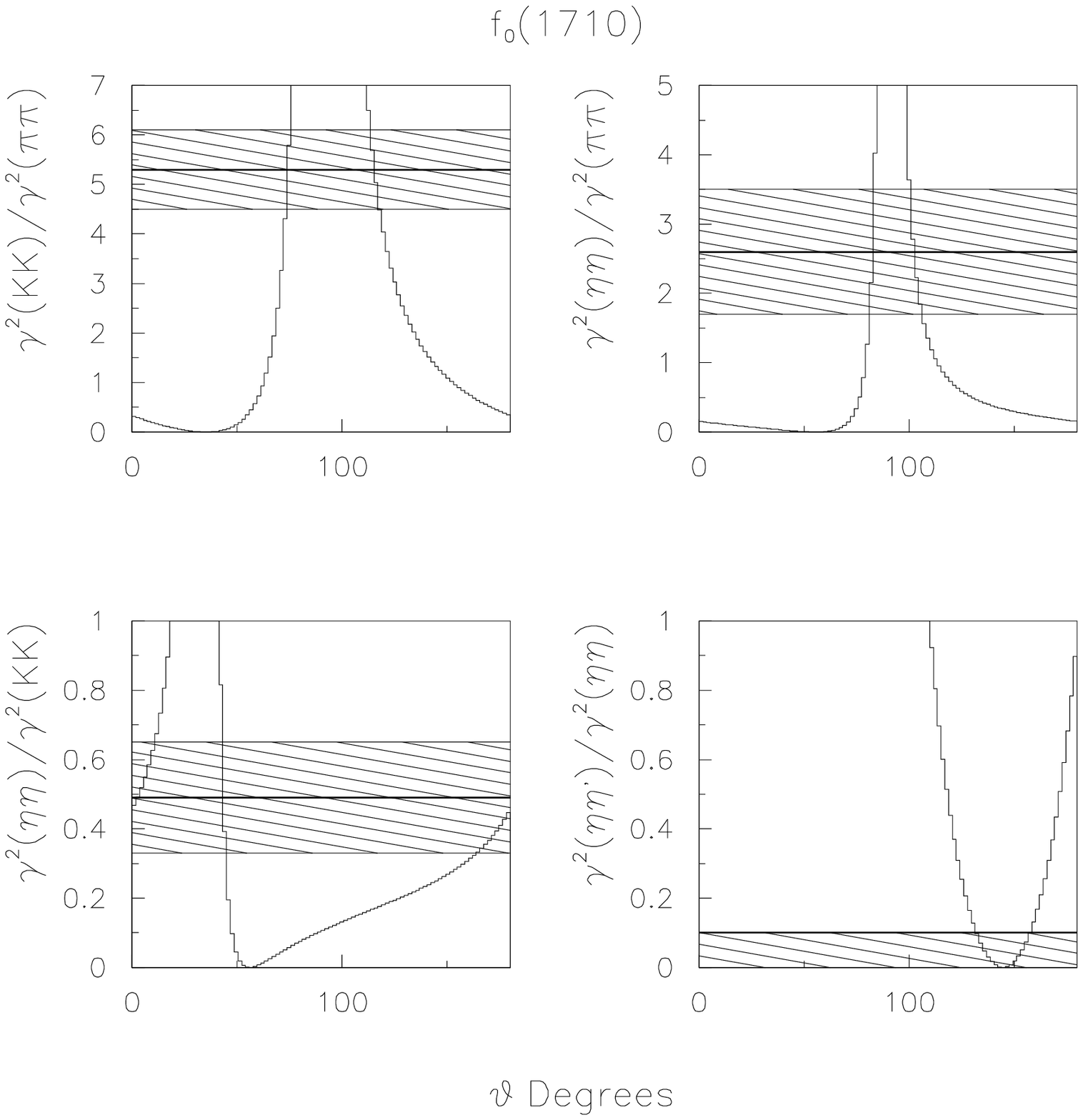,height=22cm,width=17cm}
\end{center}
\begin{center} {Figure 4} \end{center}
\end{document}